\documentstyle[12pt,graphicx,subfigure]{article}
\def\beq{\begin{equation}}
\def\eeq{\end{equation}}
\def\lsim{\ ^<\llap{$_\sim$}\ }
\def\gsim{\ ^>\llap{$_\sim$}\ }
\def\r2{\sqrt 2}
\def\beq{\begin{equation}}
\def\eeq{\end{equation}}
\def\beqn{\begin{eqnarray}}
\def\eeqn{\end{eqnarray}}

\def\sinW2{\sin^2\theta_W}

\def\mz2{M_{z}^2}
\def\c2b{\cos 2\beta}

\def\mz{M_z}

\def\Fq2{F_{2}(q^2)}

\def\sec2w{sec^2\theta_W}

\def\gmin2{(g-2)_\mu}

\catcode`\@=11 

\def\lsim{\mathrel{\mathpalette\@versim<}}
\def\gsim{\mathrel{\mathpalette\@versim>}}
\def\@versim#1#2{\vcenter{\offinterlineskip
    \ialign{$\m@th#1\hfil##\hfil$\crcr#2\crcr\sim\crcr } }}

\begin{document}
\begin{flushright}
\end{flushright}
\begin{center}
{\Large\bf Modular Invariance, Soft Breaking, $\mu$ and
$\tan\beta$ in Superstring Models\\} \vglue 0.5cm {  Pran Nath and
Tomasz R. Taylor \vglue 0.2cm {\em Department of Physics,
Northeastern University, Boston, MA 02115, USA\\} }
\end{center}
\begin{abstract}
We go beyond parameterizations of soft terms in superstring
models and investigate the dynamical assumptions that lead to the
relative strength of the dilaton {\it vs} the moduli contributions
in the soft breaking. Specifically, we discuss in some simple
heterotic orbifold models sufficient conditions to achieve
dilaton dominance.
Assuming self-dual points to be minima we find 
multiple solutions to the trilinear
and bilinear soft parameters $A_0$ and $B_0$. We discuss the
constraints on $\mu$ and $\tan\beta$ in superstring models in the
context of radiative breaking of the electroweak symmetry. We
show that string models prefer a small to a moderate value of
$\tan\beta$, {\it i.e}.\ $\tan\beta \leq 10$, and a value much larger
than this requires a high degree of fine tuning. Further, we show
that for large $\tan\beta$ the radiative electroweak symmetry
breaking constraint leads to a value
$\alpha_{string}=g_{string}^2/4\pi$ which is typically an order of
magnitude smaller than implied by the LEP data and the heterotic
superstring relation $g_{string}=k_ig_i$, where $g_i$ is the gauge
coupling constant for the gauge group ${\cal G}_i$ and $k_i$ is the
corresponding Kac-Moody level in the class of models considered.
This situation can be overcome by another fine tuned 
cancellation between the dilaton and the moduli contributions in the
soft parameters.
\end{abstract}

One of the challenges facing string theory is to generate  a
unified model of interactions which includes in it the successes
of the standard model. Many attempts have been made over the
years in this direction. This includes model building within the
heterotic string framework, {\it i.e}.\ models based on
Calabi-Yau compactifications and orbifolds \cite{heterotic}, and
models based on M-theory and D-branes \cite{dbrane}. In this paper
we examine soft breaking  in some simple heterotic models, under
the constraints of modular invariance ($T$-duality), and
investigate the dynamical conditions that govern the relative
strength of the dilaton and the moduli contribution to the soft
parameters. We also discuss the constraints that relate $\mu$ and
$\tan\beta$ in string theory in the context of radiative breaking
of the electroweak symmetry.

The scalar potential in supergravity and string theory is given
by \cite{can,cremmer} 
\beqn
V=e^{d}[(d^{-1})^i_j D_iW D_j^{\dagger}W^{\dagger}
 -3 WW^{\dagger}] +V_{D-term},
\eeqn 
where $d$ is the K\"ahler potential, $W$ is superpotential  and
 $D_iW=W_i + d_iW$, with the subscripts denoting
derivatives w.r.t.\ to the correspon\-ding fields. We will focus
our attention on the heterotic superstring compactifications on
orbifolds, although without going into their details. The only
constraint that we want to use is the $T$-duality symmetry, from
which we pick up a generic SL(2,Z) subgroup of modular invariance
associated to large--small radius symmetry. Specifically, the
scalar potential in the effective four dimensional theory depends
on the  dilaton field $S$ and on the (K\"ahler) moduli fields
$T_i$,\footnote{For simplicity, we do not discuss here the
dependence on the (complex structure) $U$-moduli.} and it is
invariant under the modular transformations \beqn T_i\rightarrow
T'_i=\frac{a_iT_i-ib_i}{ic_iT_i+d_i},
 (a_id_i-b_ic_i)=1, ~~~a_i,b_i,c_i,d_i \in Z
\eeqn Under the modular transformations, $d$ and $W$ undergo a
K\"ahler transformation: 
$W{\rightarrow}We^{-f}$,
$d\rightarrow d+f+\bar{f}$, $f=\sum_ilog(ic_iT_i+d_i)$ 
while the  combination $G\equiv d+ln(WW^{\dagger})$ is invariant.  Further,
in general, if a  function $f$ transforms under  modular
transformations as $f\rightarrow (ic_iT_i+d_i)^{n_1}$ $(-ic_i\bar
T+d_i)^{n_2}f$ then it is assigned the weight $(n_1,n_2)$. The
constraints of duality have proven useful in the investigation of
gaugino condensation and SUSY breaking in previous analyses
\cite{fmtv,brignole}. In our analysis  we will assume that $W$ is
decomposable  as $W=W_h + W_v$, where $W_h$ is the superpotential
which depends only on the fields of the hidden sector and
 $W_v$ is the superpotential for the physical fields,
  {\it i.e}.\ quarks,
 leptons and Higgs fields.
 The origin of supersymmetry breaking in string theory is not yet fully
 understood. However, one conjectures that it originates in the
 hidden sector via gaugino condensation \cite{nilles,taylor}.
We will not address this issue here but assume that stable minima
exist and supersymmetry breaking can be achieved. We are
interested in the nature of the soft terms that  appear and the
constraints on them from radiative breaking of the electroweak
symmetry. We will discuss some specific  models based on the
generic form of the K\"ahler potential and of the superpotential.
Thus for the K\"ahler potential we assume: \beqn d= D -\sum_i
log(T_i+\bar T_i) + K_{IJ} Q_I^{\dagger}Q_J +H_{IJ} Q_IQ_J, \eeqn
where, as a model for $D$, one may consider
 $D= -ln(S+\bar S +\frac{1}{4\pi^2} \sum_i^3 \delta_i^{GS}
 log(T_i+\bar T_i))$.
Here, $\delta_i^{GS}$ is the one  loop correction to the K\"ahler
potential from the Green-Schwarz mechanism \cite{antaylor}. For
the superpotential $W_v$ we assume a form  $W_v= \tilde \mu_{IJ}
Q_IQ_J + \lambda_{IJK}Q_IQ_JQ_K$, 
 where $Q$ are the matter fields consisting of the quarks, leptons
and the Higgs. Under $T$-duality, $Q$'s transform as
$Q_I\rightarrow Q_I\Pi_i (ic_iT_i+d_i)^{n^i_{Q_I}}$.
In general, $K_{IJ}, H_{IJ}, \mu_{IJ}$ and $\lambda_{IJK}$ are
functions of the moduli. The constraints on $n^i_{Q_I}$ are such
that $G$ is modular invariant. 

Soft breaking in string models has
been studied by many authors. However, most of these analyses
have been at the level of parameterizations of the breaking. We
are interested in investigating more deeply the dynamical
underpinnings of soft breaking in string models, specifically in
determining the dynamical constraints needed to achieve dilaton
dominance or admixtures of dilaton and moduli participation in
the breaking. 
Further, modular invariance  implies that the scalar potential is
stationary at the self dual points. The exact nature of these 
stationary points depends on detailed dynamical considerations
which do not address here and for the purpose of this analysis we 
assume that the self dual points are indeed minima. The conclusions
of this analysis  would be essentially unaffected if the true minima
were not exactly at the self dual points.
Our focus will be the Higgs sector of the theory since it
is this sector that controls the electroweak symmetry breaking
and much of the low energy physics of sparticles that will be
hunted at the particle accelerators. To keep the analysis simple
we impose the tree level condition $\delta^{GS}_i=0$. We also make
the simplifying assumptions that $H_{IJ}=0$.
These assumptions would
not necessarily hold in a realistic string model but some of the
lessons of the analysis may be helpful in string model building.
Below we consider three models in their increasing level of
complexity.
 
The first model we consider is where the set of moduli
fields is limited to the dilaton $S$ and one overall modulus $T$.
We consider a K\"ahler potential of the form
\beqn d=D(S,\bar S) -3ln(T+\bar T) +  h_1^{\dagger}h_1 +  h_2^{\dagger}h_2
+\sum_{\alpha \neq h_1,h_2}
(T+\bar T)^{n_{c_{\alpha}}} C_{\alpha}^{\dagger}C_{\alpha}
 \eeqn
 where $h_{1,2}$ are the two higgs doublets of the  minimal 
 supersymmetric standard model (MSSM) with  modular weights (0,0) and 
 modular
 invariance implies that the remaining  fields of MSSM obey the condition 
 $n_Q+n_{U^c}=n_Q+n_{D^c}=n_L+n_{E^c}=-3$.  
One of the constraints on the scalar potential is that of the vanishing
of the vacuum energy which in this case requires
 \beqn
|\gamma_S|^2  +|\gamma_T|^2 =1,~~ |\gamma_S|^2\equiv -\frac{1}{3}\left.
(d^{-1})^S_{\bar S} \frac{D_SW D_{\bar 
S}{W^{\dagger}}}{WW^{\dagger}}\right|_0,~~ |\gamma_T|^2\equiv
-\frac{1}{3} \left. (d^{-1})^T_{\bar T} \frac{D_TW D_{\bar 
T}{W^{\dagger}}}{WW^{\dagger}}\right|_0 \eeqn where
the subscript 0 means that we are evaluating the quantities in the
vacuum state. Now for the model of Eq.(4) we find 
$|\gamma_T|^2= \frac{(T+\bar T)^2}{9} \frac{D_TW_hD_{\bar T}
W^{\dagger}_h}{W_hW^{\dagger}_h}$, with $D_TW_h =\partial_T W_h
-\frac{3W_h}{T+\bar T}$.  
We assume that the modular dependence of $W_h$ is of the form 
\beqn
W_h=\frac{F(S,T)}{\eta(T)^{6}},
\eeqn 
where $\eta(T)$ is the
Dedekind function and $F(S,T)$ is modular  invariant, in general
a function of the absolute modular invariant $j(T)$
\cite{chandra}. Under
this assumption, one finds 
\beqn |\gamma_T|^2= (T+\bar T)^2
(G_2(T)-\frac{1}{3}D_TlogF(S,T) ) (G_2(\bar T) -\frac{1}{3} D_{\bar
T}log \bar F(\bar S,\bar T) ) 
\eeqn 
where $G_2(T)\equiv 2\partial_T ln \eta(T)+1/{(T+\bar T)}$.
The modular invariance of
$|\gamma_T|^2$ is easily checked from the transformation
properties of $\eta(T)$ and $\tilde G_2=G_2-
\frac{1}{3}D_TlogF(S,T)$, {\it i.e}.\
$ \eta (T)\rightarrow
(icT+d)^{1/2} \eta (T)$, ~~$\tilde G_2(S,T)\rightarrow (icT+d)^{2}$
$\tilde G_2(S,T)$.  For MSSM, $W_v$  takes on the form 
$W_v=\mu h_1h_2$ +
$\sum_{\alpha\beta\gamma} w_{\alpha\beta\gamma}$,
where $w_{\alpha\beta\gamma}=\lambda_{\alpha\beta\gamma} C_{\alpha}
C_{\beta} C_{\gamma}$
 are the cubic interactions invariant
under $SU(3)_C\times SU(2)_L\times U(1)_Y$ and  $\lambda_{\alpha\beta\gamma}$
are the Yukawas. Following the usual technique of computation
of soft terms \cite{can,bfs,hlw,il}, one gets 
\beqn
V_{(Soft)}\!\!&=&m_{3/2}^2((h_1^{\dagger}h_1 + h_2^{\dagger}h_2) +\!\!
\sum_{\alpha \neq h_1,h_2} (T+\bar T)^{n_{C_{\alpha}}}C_{\alpha}^{\dagger}
C_{\alpha})
+(B_0\mu h_1h_2 + \sum_{\alpha\beta\gamma} 
A_{\alpha\beta\gamma}^0w_{\alpha\beta\gamma}+ H.c.),\nonumber\\ 
m_{3/2}^2 &=& \langle e^{-G}\rangle=\langle\frac{e^DW_hW_h^{\dagger}}{(T+\bar 
T)^3}\rangle,\nonumber\\
A_{\alpha\beta\gamma}^0&=& \bar m (3|\gamma_T|^2 -\sqrt 3
|\gamma_S|(1-(S+\bar S)\partial_S ln\lambda_{\alpha\beta\gamma}) 
e^{-i\theta_s}),\nonumber\\
 B_0&=& \bar m (-1+3|\gamma_T|^2 -\sqrt
3 |\gamma_S|(1-(S+\bar S)\partial_S ln\tilde\mu) e^{-i\theta_s}),
\eeqn 
with $(S+\bar S)G_{,S}/\sqrt
3=|\gamma_S|e^{i\theta_s}$, $\bar m= m_{3/2}
\frac{e^{D/2-i\theta}}{(T+\bar T)^{3/2}}$, and where
$\langle W_h\rangle/|\langle W_h\rangle |=e^{i\theta}$. Noting that 
$W_hW_h^{\dagger}$ has
modular weight (-3,-3) while $T + \bar T$ has modular weight
(-1,-1), one finds that $ m_{3/2}^2 $ is modular  invariant.
Further, $\mu h_1h_2$  and $w_{\alpha\beta\gamma}$
 in Eq.(8) have modular weight (-3,0) each while $A^0$
and $B_0$ have modular weights (3,0) each, and thus $V_{(Soft)}$
is explicitly modular invariant. 
We note that we cannot go to the canonical basis, where the kinetic energies
of all the fields (including quarks and leptons) are normalized, at arbitrary points in the
moduli space  without destroying the holomorphicity
of the superpotential. However, we can do so once the moduli are
fixed, such as by going to the self dual points $T=1, e^{i\pi/6}$. 
where we assume the potential is minimized. 
 Here $\bar m$ takes on the values 
 $\bar m=m_{3/2}(\frac{e^{D/2-i\theta}}{2\sqrt 2},
\frac{e^{D/2-i\theta}}{3^{3/4}})$ for $T=(1,e^{i\pi/6})$.
We distinguish now the following two cases:
(i) $F$ has a non-trivial $T$-dependence: Here the vanishing of
$G_2(T)$ at the self dual points gives 
$|\gamma_T|^2=
\frac{1}{9}(T+\bar T)^2 D_Tlog F(S,T) D_{\bar T}log \bar F(\bar S,\bar T)$.
In this case one has both dilaton and moduli participation
in the soft  breaking  at the self dual
points.
(ii) $F$ has no dependence on $T$: Here the vanishing of $G_2(T)$
at the self dual points gives 
$ \gamma_T =0$, $|\gamma_S|=1$,
 for $T=1, e^{i\pi/6}$ and leads to dilaton dominance of
soft breaking at the self dual points. 
We normalize the quark and lepton fields and denote the normalized 
fields by lower case symbols, 
$c_{\alpha}=q,u^c,d^c,l, e^c$ and 
denote the Yukawas for the normalized fields by $Y_{\alpha\beta\gamma}$
so that
$w_{\alpha\beta\gamma}=\lambda_{\alpha\beta\gamma}C_{\alpha}C_{\beta}C_{\gamma}$=
$Y_{\alpha\beta\gamma} c_{\alpha}c_{\beta}c_{\gamma}$.
Further, we limit ourselves to the case where $\mu$ and 
$Y_{\alpha\beta\gamma}$ have no dependence on the dilaton field
so that
$A_0=A_{\alpha\beta\gamma}^0$. Then $A_0,B_0$ take on the
following values at the self-dual points: 
\beqn (A_0,B_0)=\left(\matrix{ -\frac{\sqrt
3}{2\sqrt 2} m_{3/2} e^{D/2-i(\theta+\theta_s)}, ~~
\frac{e^{D/2-i\theta}}{2\sqrt 2} m_{3/2} (-1-\sqrt 3
e^{-i\theta_s})\cr -\frac{1}{3^{1/4}} m_{3/2}
e^{D/2-i(\theta+\theta_s)}, ~~ \frac{e^{D/2-i\theta}}{3^{3/4}}
m_{3/2} (-1-\sqrt 3 e^{-i\theta_s})}\right) ~~~\left(\matrix{
T=1\cr T=e^{i\pi/6} }\right) 
\eeqn 

Next we consider a  model where the K\"ahler potential is similar
to that of Eq.(4), except that the  Higgs fields also have modular
weights. Thus we consider a K\"ahler potential of the following
form \beqn d=D(S,\bar S) -3ln(T+\bar T) + 
\sum_{\alpha}(T+\bar T)^{n_{C_{\alpha}}} C_{\alpha}^{\dagger}C_{\alpha}
  \eeqn
where the sum on $\alpha$ now runs over all the MSSM fields.
The vanishing of the vacuum energy again gives Eq.(5) and
a computation of the soft terms  gives
\beqn
V_{(Soft)}=m_{3/2}^2
( \sum_{\alpha}(T+\bar T)^{n_{C_{\alpha}}} C_{\alpha}^{\dagger}C_{\alpha}) 
+ (B_0\tilde\mu  H_1H_2 + 
\sum_{\alpha\beta\gamma} 
A_{\alpha\beta\gamma}^0w_{\alpha\beta\gamma}+ H.c.)
\eeqn
where $m_{3/2}$, $A_{\alpha\beta\gamma}^0, B_0$ are
again easily computed as in the preceding case. The modular invariance of
$V_{soft}$ is easily checked using $n_Q+n_{H_2}+n_{U^c}=-3$,
$n_Q+n_{H_1}+n_{D^c}=-3$, $n_L+n_{H_1}+n_{e^c}=-3$.
We again consider the case 
where $\mu$ and $\lambda_{\alpha\beta\gamma}$ have no dilaton
dependence. As in the above example, we cannot normalize the fields
at arbitrary points in the moduli space but can do so at the self dual
points. 
 Going to  the basis where the fields are canonically normalized 
 at the self dual  points we can write 
$V_{(Soft)}$=$m_{3/2}^2( \sum_{\alpha}c_{\alpha}^{\dagger}c_{\alpha}$$ )
+ $$ (B_0 W^{(2)} +A_0 W^{(3)} +H.c.)$
 where  $W^{(2)} = \mu h_1h_2$, and
$W^{(3)}$= $Y_{qh_2u^c} qh_2u^c$
+ $Y_{qh_1d^c} qh_1d^c$ + $Y_{lh_1e^c} lh_1e^c$.
Here $h_1, h_2, q, u^c, d^c,l,e^c$ are the normalized fields and the
factors needed to normalize the fields  have been absorbed in
$\mu$, $Y_{qh_2u^c}$ etc so that
$\mu=\tilde\mu/[(T+\bar T)_{SD}^{n_{H_1}/2} 
(T+\bar T)_{SD}^{n_{H_2}/2}]$ and
$Y_{qh_2u^c}$ = $\lambda_{qh_2u^c}$
$/[(T+\bar T)_{SD}^{n_{H_2}/2}$ $(T+\bar T)_{SD}^{n_Q/2}$
$(T+\bar T)_{SD}^{n_{U^c}/2}]$ ~etc.

Finally, we consider the model with many moduli.
 We take for our K\"ahler potential the form
\beqn
d=D(S,\bar S) -\sum_iln(T_i+\bar T_i) +
\sum_{i\alpha}(T_i+\bar T_i)^{n_{C_{\alpha}}^i} C_{\alpha}^{\dagger}C_{\alpha}
 \eeqn
Here, for each $T_i$, the superpotential has modular weight
-1 and one has
$D_{T_i}(W) = \partial_{T_i} W-\frac{1}{T_i+\bar T_i}W$, and
$D_{T_i}(W) = - \tilde G_2(S,T_i) W$  for each i,
where
$\tilde G_2(S,T_i)= G_2(T_i) -D_{T_i}log F(S,T_i)$.
The condition for the vanishing of
the vacuum energy in this case is
$|\gamma_S|^2+ \sum_i |\gamma_i|^2 =1$
where
$|\gamma_i|^2=-\frac{1}{3}(G^{-1})^i_{\bar i}  G_i G_{\bar i}$.
An analysis similar to the one before gives
$|\gamma_i|^2$=$\frac{(T_i +\bar T_i)^2}{3}$ 
$\tilde G_2(S,T_i)$$\tilde G_2(\bar S,\bar T_i)$.
We note that $|\gamma_i|^2$ has a relative factor of $1/3$ compared
to $|\gamma_T|^2$. In this case,
\beqn
V_{soft}=m_{3/2}^2 (\sum_{i,\alpha}(T_i+\bar T_i)^{n_{C_{\alpha}}^i}
C_{\alpha}^{\dagger}C_{\alpha})
 + (B_0\tilde\mu H_1H_2 + 
\sum_{\alpha\beta\gamma} 
A_{\alpha\beta\gamma}^0w_{\alpha\beta\gamma}+ H.c.)\nonumber\\ 
A_{\alpha\beta\gamma}^0= \bar m (3\sum_i |\gamma_i|^2 -\sqrt 3 |\gamma_S|
(1-(S+\bar S)\partial_S ln\lambda_{\alpha\beta\gamma}) 
e^{-i\theta_s}),\nonumber\\
~ B_0= \bar m (-1+3\sum_i|\gamma_i|^2 -\sqrt 3
|\gamma_S| (1-(S+\bar S)\partial_S ln\tilde\mu) 
e^{-i\theta_s}),
\eeqn 
with $\bar m= m_{3/2}{e^{D/2-i\theta}}/{\Pi_i(T_i+\bar T_i)^{1/2}}$. 
In the overall modulus case $\sum_i |\gamma_i|^2=|\gamma_T|^2$
and the result of Eq.(8) can be
obtained from Eq.(13) by setting  $T_1=T_2 =T_3 =T$. 
We can go to the canonical basis as  in previous cases. Further, since
normalizing factors do not have a dilaton dependence we can replace
$\lambda_{\alpha\beta\gamma}$ by $Y_{\alpha\beta\gamma}$
and replace $\tilde \mu$ by $\mu$ in Eq.(13).
Thus  at the self dual points $T_i=(1, e^{i\pi/6})$, and $A^0$ and $B_0$ of
Eq.(13) reduce to
 \beqn A_{\alpha\beta\gamma}^0= m_{3/2} e^{D/2-i\theta}
 (3\sum_i |\gamma_i|^2 -\sqrt 3 |\gamma_S|
 (1-(S+\bar S)\partial_S lnY_{\alpha\beta\gamma}) 
  e^{-i\theta_s})/f_{\alpha}^{\frac{1}{2}}\nonumber\\
 B_0= m_{3/2} e^{D/2-i\theta}
 (-1+3\sum_i |\gamma_i|^2 -\sqrt 3 |\gamma_S|
 (1-(S+\bar S)\partial_S ln\mu) 
  e^{-i\theta_s})/f_{\alpha}^{\frac{1}{2}}
\eeqn
where   $\{f_{\alpha}\}=8,4\sqrt 3, 6,3\sqrt 3$ and the 
 multiplicity of $f_{\alpha}$ arises from the degeneracy
 of the allowed vacua.
Again if $D_{T_i}F=0$, {\it i.e}.\ if $F$ has no $T_i$ dependence
then $\gamma_i=0$  and soft breaking at the self dual points is
dilaton-dominated. However, if $D_{T_i}F$ is non vanishing at the
self dual point, then $\gamma_i$ are also non vanishing and moduli
enter in soft breaking. In actual string calculations one does
not encounter modular invariant $F$ functions which have nontrivial
$T_i$ dependence. In this circumstance one has dilaton dominance
in the class of models we are considering.

Although the $\mu$ term is supersymmetric and not a soft
parameter, the  origin of $\mu$ is most likely soft breaking.
In fact, one common mechanism for its generation is in the
K\"ahler potential where an $H_1H_2$ can arise with a
dimensionless coefficient which can be naturally O(1). This term
when transfered via a K\"ahler transformation to the
superpotential gives a $\mu$ of the same size as the soft terms
\cite{giudice}. A concrete example of this mechanism  in string
theory was given in the analysis of Ref.\cite{agnt} where it was
shown that an $H_1H_2$ term does indeed arise in the K\"ahler
potential. However, this computation was for the invariant $\bf
27\,\overline{27}$ involving a generation and an anti generation.
 Thus the result of Ref.\cite{agnt} is not directly applicable
 to the case where the Higgs are both generational. The analysis of
 two-generational Higgs is more difficult since an invariant cannot
 be formed out of two {\bf 27}'s.
{}For the purpose of the present analysis we assume that a string computation
following the technique of Ref.\cite{agnt} can be extended to
determine the $\mu$ term needed in MSSM.
  In addition to the above the soft breaking contains the gaugino masses
  which are given by
  \beq
  M_{\alpha}= \frac{1}{2} {Ref_{\alpha}}^{-1}
  e^{-G/2} f_{\alpha a}(G^{-1})^a_b G^b,
  \eeq
  where  $f_{\alpha}$ is the  gauge kinetic
  energy function  and for a gauge group ${\cal G}=\Pi_{\alpha} {\cal 
G}_{\alpha}$,
  it is given by \cite{kgauge}
  $f_{\alpha} = k_{\alpha} S + \sum\frac{1}{4\pi^2}[C({\cal G}_{\alpha})
  -\sum_I T(R^{\alpha}_{Q_I})(1+2n^i_{Q_I}) -
  2k_{\alpha} \delta^{GS}_i]log(\eta (T_i)) + \dots,$
  where $k_{\alpha}$ is the Kac-Moody level for the subgroup ${\cal 
G}_{\alpha}$.
   In our investigation below it would suffice to consider just the
  tree contribution which  yields a  universal gaugino mass
  $m_{1/2}$ for the simplest case of $k_{\alpha}=1$.
  In this case one has
  \beqn
  m_{1/2}= \sqrt 3 m_{3/2} |\gamma_S| e^{- i\theta_s}
  \eeqn
  Under the assumption that $D_TF(S,T)=0$ and when one is at the self dual
  points, $\gamma_i =0 $, $|\gamma_S|=1$ and one
  has the result for the gaugino masses in the dilaton dominance
  case.

 Radiative electroweak symmetry breaking imposes important
 constraints on string model building. However, before
  discussing the constraint of radiative breaking in string theory
let us review the situation in supergravity models first. In the
minimal supergravity grand unified models one starts out with
five parameters $m_0, m_{1/2}, A_0, B_0, \mu$ at the GUT
scale \cite{can,hlw}.
 The renormalization group effects in running the SUSY parameters with
the GUT boundary conditions to the low scale allow the $H_2$ Higgs
mass to turn tachyonic, due to its couplings to the top quark,
which triggers the breaking of the electroweak symmetry. One of
the conditions for the minimization of the potential ${\partial
V_H}/{\partial v_i}=0$ yields \cite{apw} \beq \mu^2=
-\frac{1}{2}M_Z^2+ \frac{m_{h_1}^2
-m_{h_2}^2\tan^2\beta}{\tan^2\beta -1} \eeq
 where $m^2_{h_i}$ ($i=1,2$) contain the one-loop corrections
 from the effective  potential \cite{coleman} and
 $\tan\beta \equiv \langle h_2\rangle /\langle h_1\rangle$.
 In supergravity models $\mu$ is a free parameter
  and thus one uses the radiative symmetry breaking constraint to
  determine $\mu$ from Eq.(17) (see {\it e.g}.\ Ref.\cite{an}).
We discuss now the situation in string theory where
$\mu$ is determined in principle (for recent papers on phenomenology
under the constraints of modular invariance see
Refs.\cite{gaillard,kane,dent}). On the other hand the radiative
symmetry breaking equation also determines  $\mu$. How can these
two determinations, one from string theory and the other from 
radiative breaking of the electroweak symmetry breaking,  
be  reconciled?
Clearly once the string determined values of $m_{h_1}, m_{h_2}$ and 
 $\mu$ are  used in the
radiative symmetry breaking constraint Eq.(17) and since $M_Z$ is 
determined from experiment, the only thing left
to determine is $\tan\beta$ and so we write Eq.(17) in the form
\beq
 \tan^2\beta  =\frac{\mu^2+ \frac{1}{2}M_Z^2+ m_{h_1}^2 }
 { \mu^2 +\frac{1}{2}M_Z^2+m_{h_2}^2}
\eeq 
Eq.(18) imposes a stringent constraint on string models.
Specifically we show below that Eq.(18) implies that large
$\tan\beta$, {\it i.e}.\ $ \tan\beta>10$ is disfavored in string
models as such values require a high degree of fine tuning. This
fine tuning is different from the one encountered in supergravity
models where $\mu$ can be used to define the fine tuning
\cite{ccn}. In the numerical analysis below we assume no dilaton
dependence  of $\mu$ and $Y_{\alpha\beta\gamma}$.
In Fig.1 we give a plot of $\tan\beta$ vs $\mu$ for
the scenario with dilaton dominance of soft breaking. For the large
$\tan\beta$ case $A_0$ nearly vanishes as will be discussed in
the context of Fig.2 and so we have set $A_0=0$ in the analysis
of Fig.1 (our conclusions, however, derived from Fig.1 are largely
independent of the value of $A_0$).
 We notice the sharp rise in $\tan\beta$ for values of $\tan\beta$
 greater than in the range 5-10. For values of $\tan\beta$ above this
 range  the
slope as a function of $\mu$  becomes very large.
This region thus corresponds to the region of
high fine tuning. This means that if we  want values of $\tan\beta$
greater than 5-10 we will have to fine tune our moduli with extreme
accuracy. Further, we note that this fine tuning appears to be a generic
feature  of string models independent of the details of the soft terms.
The origin of this fine tuning is easily understood since
 a large $\tan\beta$ can only arise when the denominator  in Eq.(18)
nearly vanishes. In the vicinity of the point where the denominator nearly
vanishes  the sensitivity to small changes is
magnified. Thus consider as
a  measure of sensitivity the quantity $f_i$ defined by
$f_i=|\frac{1}{\tan\beta} \frac{\partial\tan\beta}{\partial t_i}|$
where $t_i$ are the moduli on which $\tan\beta$ depends. Using Eq.(18)
one finds  that $f_i\propto  \tan^2\beta$. Thus
 $|\frac{\partial \tan\beta}{\partial t_i}|\propto\tan^3\beta$
  and for large $\tan\beta$ this behavior leads to a  high degree
 of  fine tuning to achieve a large value of $\tan\beta$ as is
 seen in Fig.1. Thus we conclude that in  string models large
 $\tan\beta$ is problematic, requiring  a large degree of fine tuning
 of the moduli.

The constraints of radiative electroweak symmetry
in string models are even more severe than discussed above.
Thus the second electroweak
symmetry breaking condition can be written in the form
\beqn
\sin 2\beta = \frac{-2\mu B}{m_{h_1}^2+m_{h_2}^2 +2\mu^2}
\eeqn
 where all parameters are evaluated at the electroweak scale.
Since the quantities in Eq.(19) are all determined
in string theory, Eq.(19) becomes a constraint on the moduli themselves.
We illustrate this constraint for the large $\tan\beta$
case. We note that $\sin 2\beta$ nearly vanishes for the case of
large $\tan\beta$  and from Eq.(19) we deduce that $B$ must nearly
vanish for the case of large $\tan\beta$.
{}From Eqs.(8) and (14) we find that this can happen either 
 if there is a large
exponential suppression of several e-folds arising from the factor
$e^{D/2}$ or if there is a cancellation between the dilaton and the
moduli contributions which requires a fine tuning. We will see that
in most of the parameter space of the moduli the cancellation 
does not provide a sufficient suppression and one needs an exponential
suppression from the dilaton factor $e^{D/2}$. 
 The same exponential suppression also suppresses $A_0$.
 Now $D$ is related to the string coupling constant
\beqn
e^{-D}=\frac{2}{g_{string}^2}
\eeqn
where $g_{string}= k_ig_i$ and $k_i$ is the Kac-Moody level of the
gauge group ${\cal G}_i$ and $g_i$ is the corresponding gauge coupling
constant.
Eq.(20), upon using the value of $D$ implied by  Eq.(19), allows one
to compute  $\alpha_{string}$ at the unification scale $M_X$  in
terms of the parameters at the electroweak scale. Thus for the
model of Eq.(14), setting $\theta=\theta_s=0$,  one gets 
\beqn \alpha_{string}=
\frac{1}{2\pi} \frac{(\mu^2+ \frac{1}{2} M_Z^2 +m_{h_1}^2)^2} 
{\mu^2 m_{3/2}^2 r_B^2 (1+\epsilon_Z)^2\tan^2\beta}
\frac{f_{\alpha}}{(-1+3\sum_i |\gamma_i|^2-\sqrt 3|\gamma_S|
(1-(S+\bar S)\partial_S ln\mu))^2}
\eeqn 
where $\epsilon_Z= M_Z^2/(m_{h_1}^2 +\mu^2 -\frac{1}{2}M_Z^2
+(m_{h_1}^2 +\mu^2 +\frac{1}{2}M_Z^2)\cot^2\beta)$ and
 $r_B$ is the renormalization group coefficient that relates
$B_0$ at the unification scale $M_X$ to $B$ at the electroweak
scale, {\it i.e}.\ $B=r_B B_0$. In Fig.2 we give a plot of
$\alpha_{string}(M_X)$ at the unification scale as a function of
$\tan\beta$ for the dilaton dominance case 
(i.e., $\sum_i |\gamma_i|^2 =0$) 
 where we assume no dilaton dependence of $\mu$ and $Y_{\alpha\beta\gamma}$.. 
One finds that for large $\tan\beta$ the value  of
$\alpha_{string}(M_X)$ is far too small to be consistent with the
LEP constraints on $\alpha_{string}(M_X)$ necessary for
unification of gauge coupling constants.\footnote{See {\it e.g}.\ 
Ref.\cite{dienes}
for a review of gauge coupling unification.}
In Fig.3 we give a plot of $\alpha_{string}$ as a function of 
$\gamma_S$ for the case $\tan\beta =50$. One finds that $\alpha_{string}$
is typically small for  much of the range of $\gamma_S$ except for
 a small element where the denominator in Eq.(21) passes through
a zero. We note that compatibility with LEP data in this case can occur only
over a minuscule range at two points where the horizontal curve
intersects the vertical lines because of the rapid slope of the curves
there. Further agreement with LEP date requires a significant
cancellation between the dilaton and the moduli contributions in the
denominator in Eq.(21). In the above we assumed $k_i=1$. For the 
Kac-Moody levels $k_i>1$  the situation is even worse.
 Thus we conclude
that on the basis of the fine tuning problem and the problem of
too small a value of $\alpha_{string}$ encountered for the case
of large $\tan\beta$ that large $\tan\beta$ values are not
preferred in string models of the type we are considering. There
are important implications of this constraint for accelerator and
dark matter experiments. Thus, for example, the decay
$B^0_s\rightarrow \l^+l^-$ which requires a large $\tan\beta$ to
become visible within the sensitivities that would be achievable
at RUNII of the Tevatron \cite{in} would not have a chance of
being seen in string models unless fine tuning is  invoked.
Similarly, detection rates for the direct detection of dark matter
depend strongly on $\tan\beta$ and increase with increasing
$\tan\beta$
 and thus a small  $\tan\beta$ value would  make the search
for dark matter more difficult. On the plus side a smaller $\tan\beta$
 leads to a longer proton life time in
supersymmetric unified theories and is thus preferable from the
point of view of proton stability \cite{ellis}. The current
experimental data from LEP and from the Tevatron only put mild
lower limit constraints on $\tan\beta$ which are consistent with
the constraints on $\tan\beta$ from strings. Similarly, the
recent data from Brookhaven \cite{bnl} on g-2 gives  a difference
between experiment and theory of about $1.6\sigma$ to
$2.6\sigma$. Such a difference can be understood within string
models of the type discussed above with a value of $\tan\beta$
below 10 \cite{cn}.

In conclusion, we have investigated soft breaking in string models
under the constraints of modular invariance and additional simplifying
assumptions to understand more clearly the relationship of the dilaton and
the moduli in soft breaking. In our analysis we found sufficient
dynamical constraints that allow for dilaton dominance of soft breaking
at the self dual points.
In our analysis we assumed that the minima are at the self-dual points.
However, the constraints of modular invariance require only that 
the self dual points be either minima, maxima or saddle points, and do not 
exclude existence of other stationary points.
If the minima were away from the self dual points, the
values of $f_{\alpha}$ would be somewhat  different.
However, if they lie close to the 
self dual points, as in Ref.\cite{bailin}, the  modifications of $f_{\alpha}$ 
would be small.
Thus while the above results were derived within some model examples, it 
appears likely that they may be valid for a larger class of string models.

This work is  supported in part by NSF grant PHY-9901057. T.R.T.\ is grateful 
to Dieter L\"ust and the Institute of Physics at Humboldt University in Berlin 
for their kind hospitality during completion of this work. 
We thank Thomas Dent for bringing to our attention Ref.\cite{bailin}.

\begin{figure}
\hspace*{-0.6in}
\centering
\includegraphics[width=9cm,height=6cm]{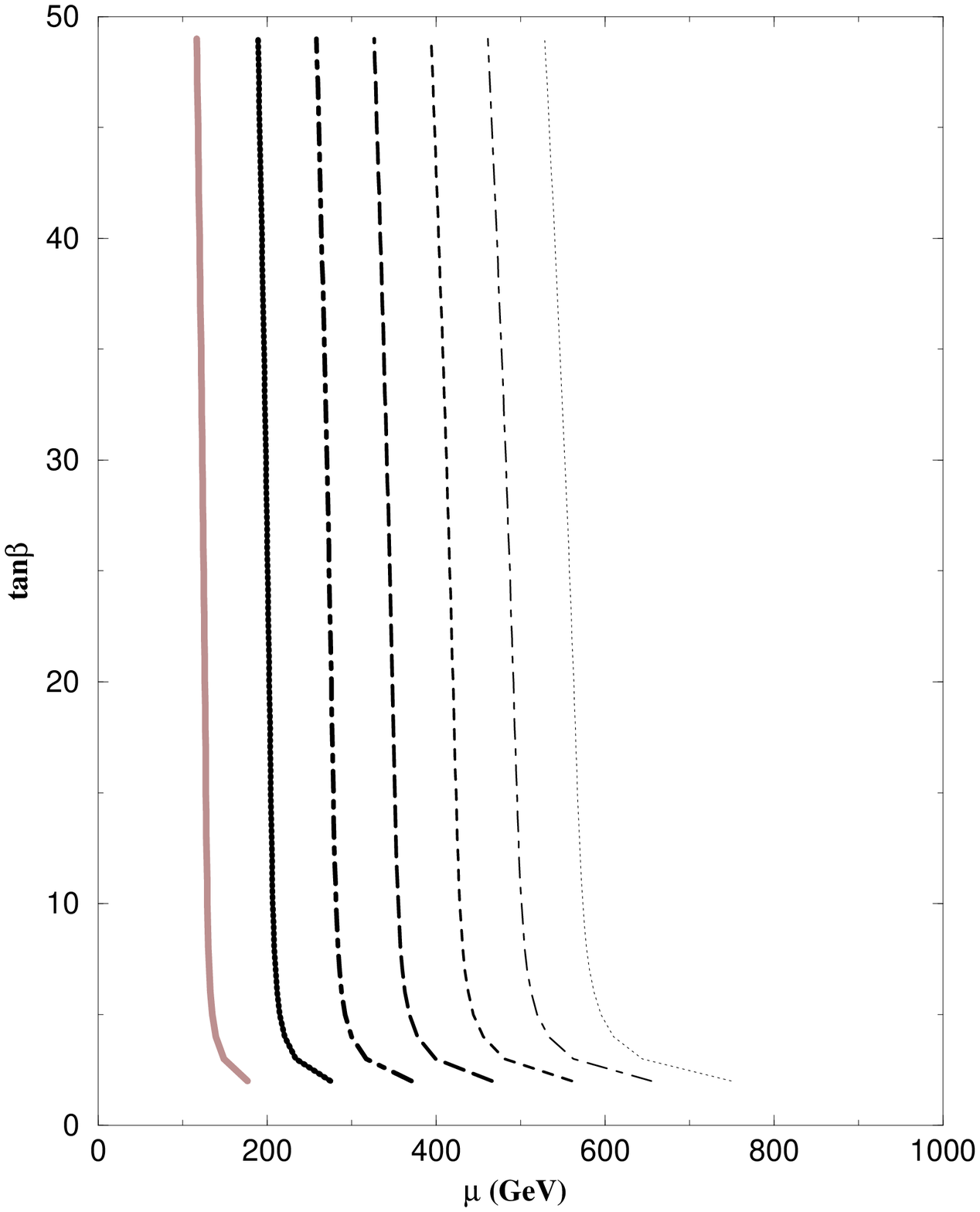}
\caption{\small Exhibition of the fine tuning problem in string models
at large tanbeta. The curves are for the scenario with dilaton
dominance of soft breaking when $m_{3/2}$ takes on the values
$50,75,100,125,150,175,200$ all masses in GeV. The Higgs mass
parameters
at the unification  scale are universal so that
$m_{h_1}(0)=m_{3/2}= m_{h_2}(0)$. }
\label{fig1}
\end{figure}
 \begin{figure}
\hspace*{-0.6in}
\centering
\includegraphics[width=9cm,height=6cm]{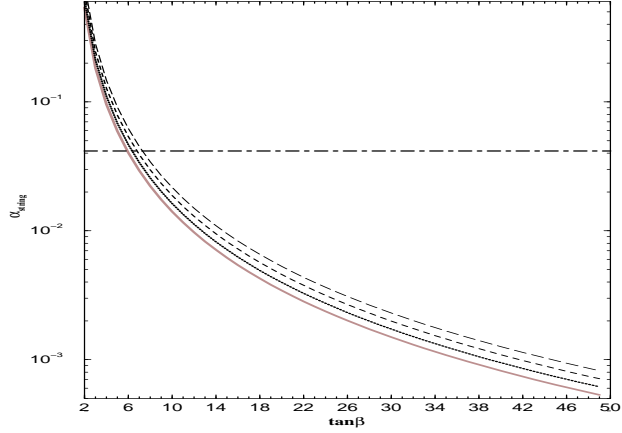}
\caption{\small Exhibition of $\alpha_{string}$ evaluated at the unification 
scale
as a function of  $\tan\beta$ for the scenario with dilaton
dominance of soft breaking when $m_{3/2}=150$ GeV
 for the case of Eq.(14) of model 3.
 The Higgs mass parameters at the unification  scale are universal so that
$m_{h_1}(0)=m_{3/2}= m_{h_2}(0)$ and $A_0=0$ as in Fig.1.
The four curves in descending order correspond to the four
degenerate vacua at the self dual points corresponding to
$f_{\alpha}=8$, $4\sqrt 3$, $6$, $3 \sqrt 3$.  The horizontal line
 is the value
of $\alpha_{string}$ at the unification scale needed for consistency with
the LEP data.}
\label{fig2}
\end{figure}
\begin{figure}
\hspace*{-0.6in}
\centering
\includegraphics[width=9cm,height=6cm]{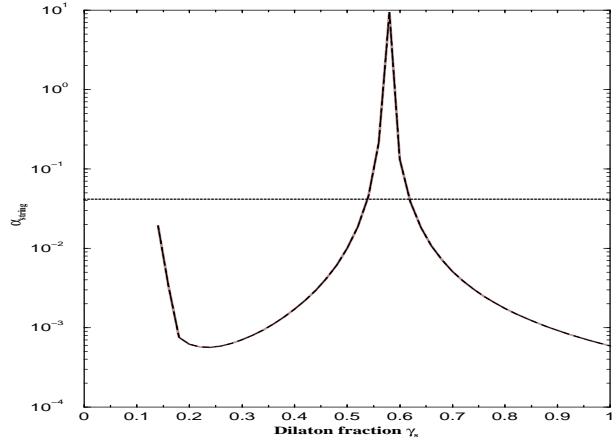}
\caption{\small Exhibition of $\alpha_{string}$ evaluated at the unification 
scale
as a function of  the dilaton fraction $\gamma_S$ 
 when $m_{3/2}=150$ GeV for the case of Eq.(14) of model 3
 with $\tan\beta =50$ and $f_{\alpha}=6$. Gaugino masses are given
 by Eq.(16) and the Higgs masses are $m_{3/2}$ at the unification 
 scale and $A_0=0$ as in Fig.1.
  All phases are set to zero. The horizontal line
 is as in Fig.2.}
\label{fig3}
\end{figure}

\end{document}